# Measurements of the spin-orbit interaction and Landé g factor in a pure-phase InAs nanowire double quantum dot in the Pauli spin-blockade regime

Jiyin Wang[1], Shaoyun Huang[*,1], Zijin Lei[1], Dong Pan[2], Jianhua Zhao[2], and H. Q. Xu[+,1,3]

[1] Key Laboratory for the Physics and Chemistry of Nanodevices and Department of Electronics, Peking University, Beijing 100871, China

[2] State Key Laboratory for Superlattices and Microstructures, Institute of Semiconductors, Chinese Academy of Sciences, Beijing 100083, China

[3] Division of Solid State Physics, Lund University, Box 118, S-22100 Lund, Sweden

Abstract

We demonstrate direct measurements of the spin-orbit interaction and Landé g factors in a semiconductor nanowire double quantum dot. The device is made from a single-crystal pure-phase InAs nanowire on top of an array of finger gates on a Si/SiO$_2$ substrate and the measurements are performed in the Pauli spin-blockade regime. It is found that the double quantum dot exhibits a large singlet-triplet energy splitting of $\Delta_{ST}$ ~2.3 meV, a strong spin-orbit interaction of $\Delta_{SO}$ ~140 μeV, and a large and strongly level-dependent Landé g factor of ~12.5. These results imply that single-crystal pure-phase InAs nanowires are desired semiconductor nanostructures for applications in quantum information technologies.

Corresponding authors: [+]Professor H. Q. Xu (hqxu@pku.edu.cn) and [*]Dr. Shaoyun Huang (syhuang@pku.edu.cn)

Key Laboratory for Physics and Chemistry of Nanodevices and Department of Electronics, Peking University, Beijing 100871, China

Email: hqxu@pku.edu.cn and syhuang@pku.edu.cn

Spin states of electrons in a tunneling coupled semiconductor double quantum dot (DQD) are one of pioneer systems for applications in solid-state based quantum information technologies.[1] In such a DQD, a spin qubit can be defined based on the coherent configuration of two electron spins. The high tunability on tunneling barriers, electron filling, and inter-dot coupling in the DQD is essential to allow for reliable initialization and manipulation of the spin qubit.[2,3] Various DQDs have been realized using semiconductor heterostructures.[2-9] In recent years, DQDs defined in InAs nanowires have attracted great attention,[10-12] because the nanowires possess a small electron effective mass, a large electron Landé g factor, strong spin-orbit interaction, and strong radial confinement to elelctrons.[13-15] Apparently, high crystal quality along an entire nanowire is indispensably desired to precisely define a DQD and thus a spin qubit in the nanowire. Single-crystal pure-phase InAs nanowires have been grown by molecular beam epitaxy (MBE)[16] and these nanowires manifest their high crystal quality by excellent performance in the field-effect transistors made from them.[17,18] In this letter, we report on realization of highly tunable DQD devices from single-crystal pure-phase InAs nanowires using the local finger gate technique and on measurements of the electron transport characteristics of the devices. We observe a large singlet-triplet energy splitting of ~2.3 meV, which is on demand for robust initialization of spin qubits. A large and level-dependent Landé g factor ($|g^*|$~12.5) and strong spin-orbit interaction (with an interaction energy of $\Delta_{SO}$ ~140 µeV) are also extracted from the measurements.

The InAs nanowires employed in the device fabrication for this work are grown by MBE and are of 3-5 µm in length and 10-50 nm in diameter. It has been found that the InAs nanowires grown by MBE with small diameters (< 50 nm) are in pure phase and are either single wurtzite or single zincblende crystals, free from stack faults and extended defects.[16] The device fabrication begins from preparation of finger gate arrays on a silicon substrate covered with a 200-nm thick layer of $SiO_2$. Each array contains seven finger gates labeled as G1 to G7 from left to right (see Fig. 1). Electron-beam lithography (EBL) is used to define patterns of the finger gates on PMMA resist and electron-beam evaporation (EBE) is used to deposit a 5-nm-thicklayer of titanium and then a 15-nm-thicklayer of gold. After lift-off process, finger gates with a width of 30 nm and a pitch of 80 nm are obtained. Subsequently, the finger gate arrays are covered by a layer of $HfO_2$ with a thickness of 10 nm by means of atomic layer deposition. Then, InAs nanowires grown by MEB are transferred from the growth substrate onto the finger gate arrays. The finger gate arrays each occupied with only one nanowire with a diameter of ~40 nm are selected for final device fabrication. Here, the nanowires with a diameter of ~40 nm are selected, because such nanowires are thin enough to be in a pure phase and are in the same time thick enough to avoid a significant increase in the contact resistance arising from quantum confinement.[19] The source and drain contact regions on two sides of each nanowire are defined and opened again by EBL. The contact regions are then briefly etched in diluted $(NH_4)_2S_x$ solution, in order to remove nanowire surface oxide and to achieve surface passivation, right prior to the metal deposition of 5-nm-thick titanium film and 90-nm-thick gold film for electrodes.[20,21] The devices are finally completed by lift-off process. Figure 1(a) shows a scanning electron microscope (SEM) image of a fabricated device.

The fabricated devices are first electrically characterized at room temperature. The devices with the source-drain resistance of 10-50 k$\Omega$ are selected for low temperature measurements. The low temperature measurements are performed in a dilution refrigerator. Several devices are measured. Below, we will report on the results of measurements of a representative device in the dilution refrigerator with a base temperature of 40 mK. In the device, the seven finger gates are able to work independently and the threshold voltage required to suppress the channel current completely by any one of the finger gates is in the range of -3 to -2.5 V. Note that the back gate and the uninvolved gates (G1 and G2) are grounded throughout the measurements.

Figure 2(a) shows the measured differential conductance $dI_{SD}/dV_{SD}$ of a single quantum dot (QD) as a function of source-drain voltage $V_{SD}$ and gate voltage $V_{G6}$ (charge stability diagram). Here the QD is defined using gates G5 and G7 by setting the gate voltages at $V_{G5}=$ -2.95 V and $V_{G7}=$ -2.55 V, the QD state is tuned by the voltage $V_{G6}$ applied to gate G6, and $I_{SD}$ is the source-drain current. Regular Coulomb blockade diamond are clearly observable in the charge stability diagram. The close points seen at the zero $V_{SD}$ between neighboring Coulomb diamonds indicate that the single electron transport occurs via the single QD. Conductance lines through zero-dimensional excited states are also clearly observed in the QD. From the measurements, the charging energy and quantization energy of the QD are extracted to be about 12 and 2 meV, respectively. These single-QD features can be observed in the nanowire device by applying similar voltages to any other three neighboring finger gates. As a consequence, a DQD can be formed in the InAs nanowire using any five neighboring finger gates, for instance, gates G3 to G7. Here, gates G3 and G7 are used to generate the two outer tunneling barriers, which manipulate the tunneling transparency between the source and the left dot (QD1) and between the drain and the right dot (QD2), respectively. Gate G5 is used to control the inter-dot coupling and gates G4 and G6 are used to tune the electrostatic potentials in QD1 and QD2, respectively. Figures 2(b)-2(d) show $I_{SD}$ measured for the DQD at a linear response voltage $V_{SD}$ =35 µV as a function of voltages $V_{G4}$ and $V_{G6}$ applied to gates G4 and G6 at three representative inter-dot coupling strengths. In the measurements, voltage $V_{G3}$ and $V_{G7}$ are set at -2.55 V and -2.30 V, respectively, to keep QD1 and QD2 in roughly equal couplings with the source and drain reservoirs. At $V_{G5}$ =0 V [Fig. 2(b)], no tunneling barrier is formed in the nanowire at the G5 position and, therefore, a larger single QD is defined between barriers G3 and G7, as shown in the inset of Fig. 2(b). Coulomb oscillation peaks, i.e., straight lines of high current in the figure, are observed when an electron is added into or extracted out of the QD at the degenerate point of the QD states. By pushing the voltage applied to gate G5 to a negative value of $V_{G5}=$ -1.85 V, a tunneling barrier is generated in the nanowire at the G5 position and a DQD is formed between gates G3 and G7, as shown in the inset of Fig. 2(c). Compared with that in Fig. 2(b), the high current lines are bent at certain positions where quantum levels at the two dots move close in energy [Fig. 2(c)]. The appeared large separations between kink pints imply that the DQD is in the strong inter-dot coupling regime, as depicted in the inset of Fig. 2(c). When $V_{G5}$ is set to -2.6 V, well defined honeycomb patterns are obtained [Fig. 2(d)]. In the lowest order transport process, current flow occurs only at the corners (triple points) of hexagons by elastic transport through the DQD.[22] Small but not negligible current present on the boundaries of the hexagons is caused by co-tunneling processes.[23] Now, the numbers of electrons in the two dots are well defined in each hexagon region and the DQD is in the weak coupling regime. From the hexagon region marked by a dashed square in Fig. 2(d), we can extract the capacitances between gate G4 and QD1 and between gate G6 and QD2 as $C_{G4}$= 6.3 aF and $C_{G6}$= 5.1aF. The corresponding total capacitances of QD1 and QD2 are estimated to be $C_4$= $C_{G4}/α_4$= 15.0 aF and $C_6$= $C_{G6}/α_6$= 13.8 aF, where conversion factors $α_4$ and $α_6$ are obtained from Fig. 3(a) as will be discussed later. Finally, the charging energies of QD1 and QD2 are found to be $E_{C,4}$= $e^2/C_4$= 10.7 meV and $E_{C,6}$= $e^2/C_6$= 11.6 meV, in good agreement with the value extracted for the single QD defined using gates G5 to G7 as in Fig. 2(a). The mutual capacitance between the two dots is found to be $C_m$= 3.3 aF, consistent with the weak coupling condition.

Figure 3(a) shows source-drain current $I_{SD}$ measured for the DQD in the weak coupling regime as a function of voltages $V_{G4}$ and $V_{G6}$ at a finite source-drain bias voltage of $V_{SD}$= - 4 mV. The source-drain bias of 4 mV is smaller than the charging energies of individual dots and the transport occurs dominantly via single-electron processes. At the finite source-drain voltage, each triple point extends into a triangular conducting region.[23] The sizes of the triangle, $ΔV_{G4}$ and $ΔV_{G6}$ as indicated in Fig. 3(a), are related to $V_{SD}$ via $V_{SD}$ =$α_4×ΔV_{G4}$ and $α_6×ΔV_{G6}$. Inside the triangle, electrons are energetically able to tunnel from the right dot to the left dot via the transition from the S(0,2) singlet state to the S(1,1) singlet state [S(0,2)→S(1,1)]. Here, the left (right) number in the brackets indicates the effective number of electrons in QD1 (QD2). Note that the inner core electrons are spin-paired off and can be treated as background charges. In Fig. 3(a), we observe two enhanced current lines (marked by a red solid arrow and a red dotted arrow) parallel to the base line (marked by a black arrow) of the triangle. The two enhanced current lines can be attributed to resonant transport processes via excited states. Figure 3(b) shows the results of measurements at a reversed bias voltage of $V_{SD}= + 4$ mV. It is seen that the current in the region (marked by a green solid star) between the base line and the high current line marked by a red solid arrow is dramatically suppressed. The suppression happens because the T(1,1) triplet state is not allowed to transit to the S(0,2) singlet state without flipping electron spin, as shown in the inset of Fig. 3(d) (Pauli spin blockade).[4] However, tunneling transport can occur (1) when the T(1,1) state is aligned with the S(0,2) state via the transition of the T(1,1) state to the S(1,1) state by dynamic perturbation of nuclear spins [leading to the high current line marked by the black arrow in Fig. 3(b)] and (2) when the T(1,1) state is aligned with the T(0,2) or T$^*$(0,2) state via resonant transport processes [leading the high current lines marked by the red solid arrow and the red dotted arrow in Fig. 3(b)].[11,24] As shown in the inset of Fig. 3(d), T$^*$(0,2) is a triplet state formed by two electrons with one electron occupying the ground state and the second electron occupying a second excited state in QD2. The energy difference between the base line and the high current line marked by the red solid arrow is the singlet-triplet splitting energy $\Delta_{ST}$ [to be further discussed below in Fig. 3(e)]. It can be extracted from the measurements shown in Fig. 3(b) that $\Delta_{ST}$ in the DQD is ~1.5 meV in this gate voltage region. Figure 3(c) shows the transport characteristics of the DQD in the triangle region of the gate voltages at magnetic field $B$=1 T applied perpendicular to the substrate. In this finite magnetic field, the T(0,2) triplet state splits into three states, T$_-$(0,2), T$_0$(0,2) and T$_+$(0,2). Now, $\Delta_{ST}$ is measured by the energy difference between the S(0,2) and the

6 / 15

T$_+$(0,2) state [see Fig. 3(e)]. Thus, $\Delta_{ST}$ is smaller when compared with the value at zero magnetic field due to the Zeeman effect [see Fig. 3(e)]. Figure 3(d) shows $\Delta_{ST}$ as a function of the applied magnetic field $B$. It can be seen that with increasing $B$, $\Delta_{ST}$ decreases linearly, following the equation $\Delta_{ST}(B) = \Delta_{ST}(0) - S_Z|g^*|\mu_B B$, where $\mu_B$ is the Bohr magneton and $g^*$ is the Landé g factor.[25] Here, we note that $g^*$ is in fact $g^*_{T(0,2)}$, see discussion below. A linear fit yields $|g^*|\sim 8.3$, consistent with previous reports.[26,27]

Figure 3(e) shows the energy evolutions of the relevant states discussed above, in the spin blockade regime, as a function of detuning energy $\varepsilon$ at a finite magnetic field $B$. Here we use the S(1,1) or the T(1,1) level as the energy reference. Thus, these states do not change in energy with increasing detuning $\varepsilon$. At a finite magnetic field, the T(1,1) triplet state splits into three states of T$_-$(1,1), T$_0$(1,1) and T$_+$(1,1) as marked by red dashed lines in the figure. However, the S(0,2) and T(0,2) states decrease in energy with increasing detuning $\varepsilon$. Furthermore, in the finite magnetic field, the T(0,2) triplet state splits into three states of T$_-$(0,2), T$_0$(0,2) and T$_+$(0,2) as marked by red solid lines in the figure. Since $\Delta_{ST}$ (0) represents the energy difference between the S(0,2) and T(0,2) states at $B$=0, $\Delta_{ST}$ (0) can be obtained from the difference in detuning energy between the cross point of the T$_0$(1,1) and S(0,2) states [denoted by M] and the anticross point of the T$_0$(1,1) and T$_0$(0,2) states [denoted by N] as shown in Fig. 3(e). In the finite magnetic field, $\Delta_{ST}$ (B) is however obtained from the difference in detuning energy between the cross point of the T$_+$(1,1) and S(0,2) states [denoted by O] and the anticross point of the T$_+$(1,1) and T$_+$(0,2) states [denoted by P]. It is clear that $\Delta_{ST}(B)$ decreases with increasing $B$ as we have already seen in Figs. 3(c)-3(d). In Fig. 3(e), the energy difference between points M and O represents the splitting energy of the T$_+$(1,1) state at the finite magnetic field due to the Zeeman effect, while the energy difference between points N and P represents the difference in the energy shifts of the T$_+$(0,2) and T$_+$(1,1) states at the finite magnetic field.

Figure 4(a) shows another pair of triangles in the measurements with a less number of electrons residing in the DQD. At zero magnetic field, the singlet-triplet splitting $\Delta_{ST}$ is extracted to be ~2.3 meV in this case, which is much larger than the values reported in other works.[12,27] Figure 4(b) displays $I_{SD}$ as a function of magnetic field $B$ and detuning energy $\varepsilon$, along the dashed arrow in Fig. 4(a). With increasing $B$, the base current line [T$_+$(1,1)→S(0,2)] shifts towards large detuning energy $\varepsilon$, while the high current line [T$_+$(1,1)→ T$_+$(0,2)] shifts towards small $\varepsilon$. Thus, the splitting energy $\Delta_{ST}$ reduces with increasing $B$. At $B$~3.45 T, an anti crossing of the two current lines shows up because of the spin-orbit interaction that mixes the state S(0,2) with the state T$_+$(0,2). Here we emphasize that the high current line corresponding to transition T$_+$(1,1)→ T$_+$(0,2) can also vary with $B$ because the Landé g factors $g^*_{T(1,1)}$ and $g^*_{T(0,2)}$ of the T$_+$(1,1) and T$_+$(0,2) states can be different due to the fact that the g factor in a semiconductor nanowire QD is level dependent.[28] In the particular case of the measurements shown in Fig. 4(b), this high current line shifts to lower detuning energy $\varepsilon$ with increasing $B$ as shown in Fig. 3(e). To extract the g factors $g^*_{T(1,1)}$ and $g^*_{T(0,2)}$ and the spin-orbit interaction energy $\Delta_{SO}$, the measured data shown in the dashed square of Fig. 4(b) are fitted using a simple two-level perturbation model,[27,29]

$$E_\pm = \frac{E_S + E_{T_+}}{2} \pm \sqrt{\frac{(E_S - E_{T_+})^2}{4} + \Delta_{SO}^2}, \qquad (1)$$

where $E_{T_+}(B) = E_{T_+}(0) - [|g^*_{T(0,2)}| - |g^*_{T(1,1)}|]\mu_B B$ is the detuning energy between the two triplet T$_+$(0,2) and T$_+$(1,1) states and $E_S(B) = |g^*_{T(1,1)}|\mu_B B$ is the detuning energy between the singlet S(0,2) and triplet T$_+$(1,1) states. The result of the fitting is presented by the red solid lines in Fig. 4(c). From the fitting, we extract the g factors of the two triplet states of $|g^*_{T(1,1)}|$~6.3 and $|g^*_{T(0,2)}|$~12.5 and the spin-orbit interaction energy of $\Delta_{SO}$~140 μeV. The obtained spin-orbit interaction energy is in the same order of magnitude as previously found in QDs made from InAs and InSb nanowires.[27,28] The large difference seen in the two extracted g factors provides a clear evidence that the g factor in a semiconductor nanowire DQD is also strongly state dependent. Furthermore, from the extracted g factors of the two triplet states [T(1,1) and T(0,2)], we can estimate that the difference between the g factors of the two involved neighboring single particle states in QD2 is about 12.4, manifesting again a strong level dependence of the g factor in a semiconductor QD with a strong spin-orbit interaction.

In conclusion, we have measured the spin-orbit interaction and Landé g factors in a semiconductor nanowire DQD. The device is made from a MBE-grown single-crystal pure-phase InAs nanowire on a Si/SiO$_2$ substrate. The DQD is defined and manipulated using local finger gates beneath the nanowire. Low temperature electron transport measurements demonstrate that the DQD can be tuned to desired inter-dot coupling regimes and shows the Pauli spin-blockade effect in the weak inter-dot coupling regime. It is found that the DQD exhibits a large singlet-triplet energy splitting of ~2.3 meV, a strong spin-orbit interaction of ~140 μeV, and a large and level-dependent g factor of ~12.5. These results imply that single-crystal pure-phase InAs nanowires are desired semiconductor nanostructures for applications in quantum information technologies.

**ACKNOWLEDGEMENTS**

This work is supported by the Ministry of Science and Technology of China (MOST) (Nos. 2012CB932700 and 2012CB932703), the National Natural Science Foundation of China (Nos. 91221202, 91421303, 11274021, 61504133 and 61321001), and the Specialized Research Fund for the Doctoral Program of Higher Education of China (No. 20120001120127). HQX also acknowledges financial support from the Swedish Research Council (VR).

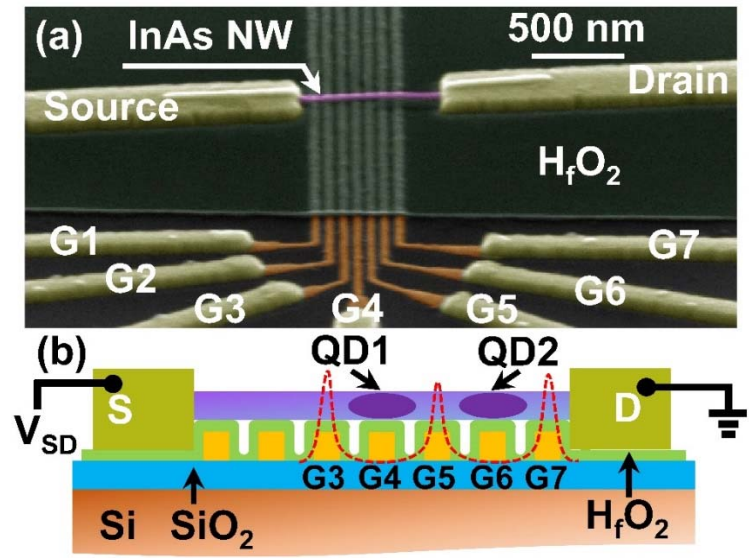

**Fig** 1. (Color online) (a) Scanning electron microscope image (false color) of a representative InAs nanowire device used in this work. The finger gates beneath the nanowire are fabricated with a width of 30 nm and a pitch of 80 nm. (b) Cross-sectional illustration of the device structure. The DQD is defined by using finger gates G3, G5 and G7 and is manipulated using finger gates G4 and G6.

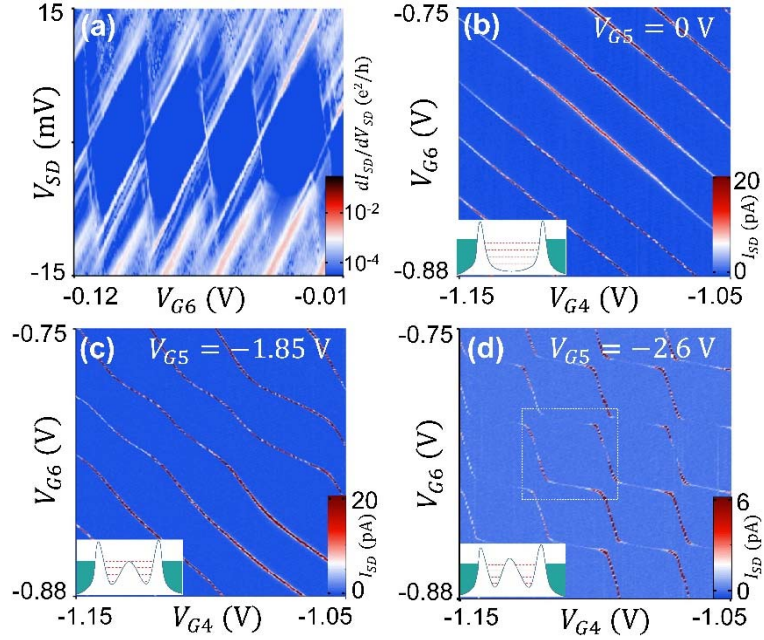

**Fig** 2. (Color online) (a) Charge stability diagram of a single QD defined using gates G5 and G7 with $V_{G5}$= -2.95 V and $V_{G7}$= -2.55 V. (b)-(d) Charge stability diagrams of the DQD defined using finger gates G3, G5 and G7 in different inter-dot tunnel coupling regimes. The voltages applied to the outer barrier gates G3 and G7 are set at $V_{G3}$= -2.55 V and $V_{G7}$= -2.3 V, and the inter-dot coupling strength is tuned by the voltage applied to gate G5. The measurements are performed at an applied source-drain bias voltage of $V_{SD}$= 35 µV. The insets in (b)-(d) show the evolution of the electrostatic potential from one big single QD to a DQD with decreasing inter-dot coupling. In (b) a large single QD is defined between gates G3 and G7 at $V_{G5}$= 0 V, in (c) the DQD with a strong inter-dot coupling is defined by setting $V_{G5}$= -1.85 V, and in (d) the DQD in the weak inter-dot coupling regime is defined by setting $V_{G5}$= -2.6 V.

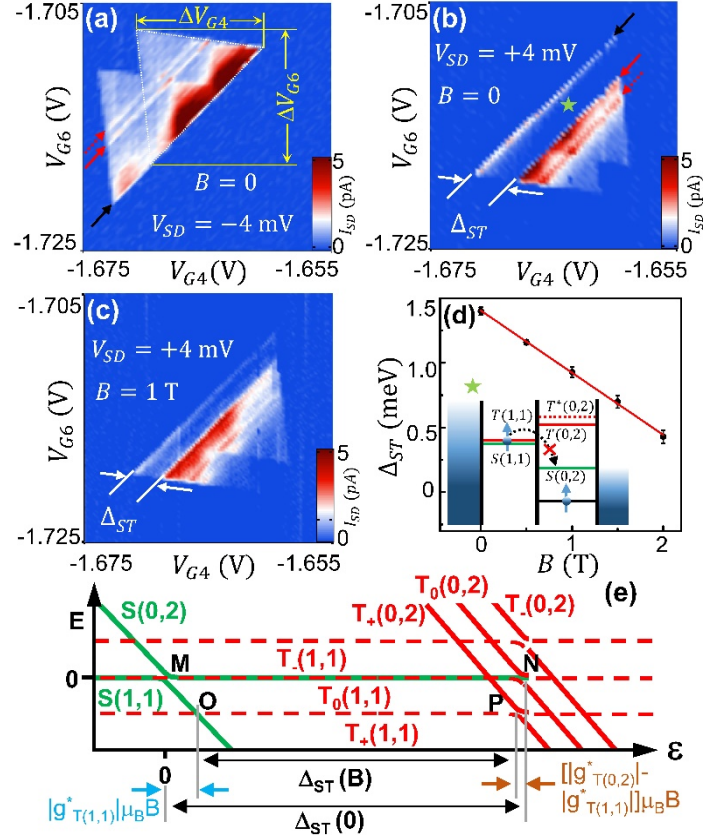

**Fig** 3. (Color online) (a) Source-drain current $I_{SD}$ measured for the DQD in the weak inter-dot coupling regime as a function of gate voltages $V_{G4}$ and $V_{G6}$ at $V_{SD}$=-4 mV and $B$=0. The DQD is defined by setting the two outer barrier gates at $V_{G3}$= -2.75 V and $V_{G7}$= -2.45 V and the inter-dot coupling gate at $V_{G5}$= -2.6 V. The black arrow and two red arrows mark resonant transport processes via the ground state and two exited states, respectively. (b) Source-drain current $I_{SD}$ measured for the DQD defined by setting the barrier and inter-dot coupling gate voltages at the same values as in (a) as a function of gate voltages $V_{G4}$ and $V_{G6}$ at a source-drain bias voltage of $V_{SD}$=+4 mV and $B$=0. The source-drain current in the region between the base line (marked by black arrow) and the high current line (marked by the red solid arrow) is dramatically suppressed due to the Pauli spin-blockade effect. (c) The same as in (b) but for a magnetic field of $B$=1 T applied perpendicular to the substrate. (d) Singlet-triplet splitting energy $\Delta_{ST}$ of the DQD as a function of magnetic field applied perpendicular to the substrate. A linear fit (the red line) yields $|g^*|$~8.3. The inset displays schematically the Pauli spin-blockade effect in the DQD. (e) Energy evolutions of different DQD states as a function of detuning energy $\varepsilon$ at a finite magnetic field $B$. Labels M, N, O and P mark the cross or anticross points between the DQD states. $\Delta_{ST}(0)$ denotes the energy difference between the S(0,2) state and the T(0,2) state at $B$=0 and $\Delta_{ST}(B)$ denotes the energy difference between the S(0,2) state and the T$_+$(0,2) state at the finite value of $B$. The energy $|g^*_{T(1,1)}|\mu_B B$ manifests the energy shift of the T$_+$(1,1) state at the finite value of $B$ with respected to that at $B$=0. The energy $[|g^*_{T(0,2)}| - |g^*_{T(1,1)}|]\mu_B B$ is determined by the difference in energy shift of the T$_+$(1,1) and T$_+$(0,2) states at the finite value of $B$.

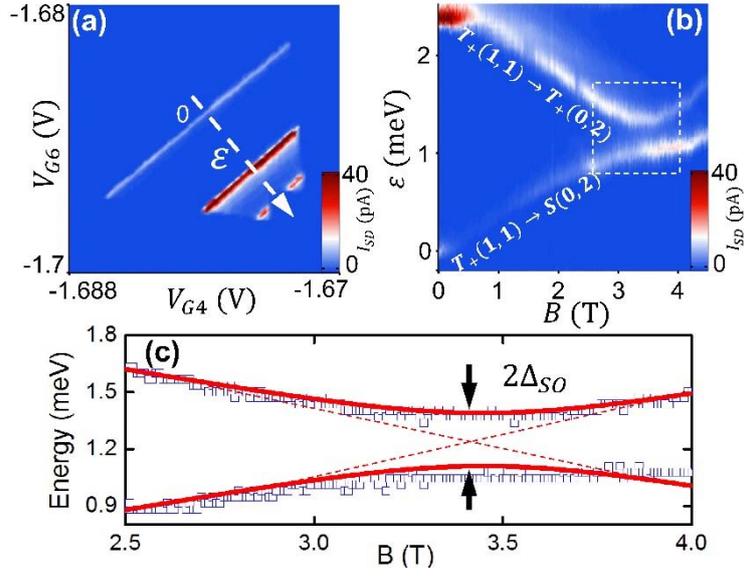

**Fig** 4. (Color online) (a) Charge stability diagram measured for the same DQD as in Fig. 3, but with a less number of electrons inside, in the Pauli spin-blockade regime. Here, a large singlet-triplet splitting $\Delta_{ST}$ is observed. (b) Source-drain current $I_{SD}$ as a function of perpendicular magnetic field $B$ and detuning energy $\varepsilon$, measured along the dashed arrow indicated in (a). (c) Measured detuning energies (blue symbols) in the region marked by a dashed square in (b) together with the results of a fit to a two-level model (red lines). Red dashed lines illustrate the behavior of the two high current lines without taking the spin-orbit interaction into account.